# $WS_2$ as a saturable absorber for ultrafast photonic applications of mode-locked and Q-switched lasers

*Kan Wu[1†,\*], Xiaoyan Zhang[2†], Jun Wang[2,\*], Xing Li[1], and Jianping Chen[1]*

† These authors contributed equally to this work.
\* Corresponding Author: E-mail: kanwu@sjtu.edu.cn, jwang@siom.ac.cn

[1]State Key Laboratory of Advanced Optical Communication Systems and Networks, Department of Electronic Engineering, Shanghai Jiao Tong University, Shanghai 200240, China
[2]Key Laboratory of Materials for High-Power Laser, Shanghai Institute of Optics and Fine Mechanics, Chinese Academy of Sciences, Shanghai 201800, China

Abstract

Two-dimensional (2D) nanomaterials, especially the transition metal sulfide semiconductors, have drawn great interests due to their potential applications in viable photonic and optoelectronic devices, such as saturable absorbers (SAs) and optical switches, etc. In this work, tungsten disulfide ($WS_2$) based SA for ultrafast photonic applications was demonstrated. $WS_2$ nanosheets were prepared using liquid-phase exfoliation method and embedded in polyvinyl alcohol (PVA) thin film for the practical usage. Saturable absorption was observed in the $WS_2$-PVA SA at the telecommunication waveband near 1550 nm. By incorporating $WS_2$-PVA SA into a fiber laser cavity, both stable mode locking operation and Q-switching operation were achieved. In the mode locking operation, the laser obtained femtosecond output pulse width and high spectral purity in the radio frequency spectrum. In the Q-switching operation, the laser had tunable repetition rate and output pulse energy of a few tens of nano joule. Our findings suggest that few-layer $WS_2$ nanosheets embedded in PVA thin film are promising nonlinear optical materials for ultrafast photonic applications as a mode locker or Q-switcher.

# 1. Introduction

Novel two-dimensional (2D) materials bring up a new area of 2D nano-systems and have attracted intense interests in the recent years. 2D semiconducting transition metal dichalcogenides (TMDs), including molybdenum disulfide ($MoS_2$) and tungsten disulfide ($WS_2$), have in particularly received significant attention because of their semiconducting property with tunable bandgaps and abundance in nature [1-4]. 2D atomically thin $WS_2$ nanosheets exfoliated from bulk counterparts have shown exotic electronic and optical properties, such as indirect-to-direct bandgap transition with reducing number of layers, high carrier mobility and strong spin-orbit coupling due to their broken inversion symmetry [5-7], which have enabled widely potential applications in viable photonic and optoelectronic devices [2, 8]. Similar to graphene, 2D $WS_2$ can be fabricated by the methods including micro-mechanical cleavage, liquid-phase exfoliation (LPE) and chemical vapor deposition [4]. Among these methods, liquid-phase preparation is scalable and permitting fabrication of wafer-scale thin films and coatings, showing good prospects for making flexible electronics and composite materials [1, 7].

In various applications of 2D materials, saturable absorbers (SAs) for mode-locked lasers and Q-switched lasers play a key role for the ultrafast photonic applications. Graphene is the first discovered 2D material and its saturable absorption has been widely investigated for the mode locked lasers and Q-switched lasers [9-11]. Graphene based mode locking and Q-switching operations have been reported for a wide range of operation wavelengths [12-16] and different incorporation methods [17, 18]. Nonlinear properties of graphene were also studied [19].

Triggered by the study of graphene, saturable absorption has been discovered in other 2D materials which opens up a door to a promising area that a wide class of 2D materials may all have saturable absorption and the choice for SAs can be significantly extended. Graphene oxide [20-23] and topological insulators [24-29] based SA have been demonstrated for their mode locking and Q-switching operations. Very recently, MoS$_2$ has been reported for its saturable absorption in a wide band from 400 nm to 2.1 μm [30-34]. Z-scan studies have revealed the high optical nonlinearity of MoS$_2$ which is comparable to that of graphene [32, 33]. Mode locking and Q-switching operations based on MoS$_2$ SA have been achieved near 1 μm, 1.55 μm and 2.1 μm [35-38]. Femtosecond pulse width down to 710 fs [39] and harmonic mode locking with a repetition rate up to 2.5 GHz [31] have also been reported. It is interesting to find that although the direct bandgap of monolayer MoS$_2$ is ~1.8eV (688 nm) and the indirect bandgap is 0.86 - 1.29 eV (1443 - 962 nm) [34], the saturable absorption property exists in a wide band beyond this limitation according to the aforementioned experimental works. This confliction may be attributed to the existence of the defects in the materials. By deliberately introducing the detects, the bandgap of MoS$_2$ has been experimentally reduced to 0.26 eV corresponding to a wavelength of 4.7 μm, indicating the potential of MoS$_2$ for ultra-broadband operation [34].

As another kind of TMD material, WS$_2$ has similar material properties compared with MoS$_2$. Therefore, it is intuitive to ask whether WS$_2$ also has saturable absorption similar to MoS$_2$? In this work, the saturable absorption of WS$_2$ nanosheets in the telecommunication waveband near 1550 nm was discovered and investigated. By embedding WS$_2$ nanosheets into PVA thin film, WS$_2$-PVA SA was obtained with a modulation depth of 2.96% and a saturation intensity of 362 MW/cm$^2$ for ultrafast photonic applications. Incorporating WS$_2$-PVA SA into a fiber laser cavity, both mode locking and Q-switching operations near 1550 nm were demonstrated. Femtosecond pulse width of 595 fs and high RF spectral purity of 75 dB extinction ratio were obtained in the mode locking operation whereas tunable repetition rate from 90 kHz to 125 kHz and pulse energy of tens of nano joule were obtained in the Q-switching operation. These findings indicate that saturable absorption indeed exists in WS$_2$ near 1550 nm and WS$_2$-PVA SA is a promising mode locker and Q switcher for the ultrafast photonic applications.

## 2. Results and discussions

### 2.1. WS$_2$-PVA saturable absorber

A high-quality WS$_2$-based saturable absorber with thin-film form provides significant flexibility for the photonic applications such as mode-locked and Q-switched fiber lasers. Liquid-phase exfoliation method was applied to prepare dispersions with large populations of monolayer and few-layer WS$_2$ using sodium cholate (SC) as surfactant. A detailed description of LPE method can be referred to the Methods section. The prepared WS$_2$/SC dispersions had a concentration of ~0.015 mg/ml. Transmission electron microscope (TEM) was utilized to confirm the existence of WS$_2$ nanosheets in the dispersions, shown in Figure 1(a). WS$_2$ nanosheets with the size of a few hundreds of nanometers can be clearly observed in the TEM image due to the effective exfoliation of the LPE method. The WS$_2$ dispersions were then mixed with PVA aqueous solution and 1-hour ultrasonic processing was performed to obtain a thorough mixture. The WS$_2$-PVA mixture was dropped on a glass plate and dried in the room temperature to form the thin film, shown in Figure 1(b). The thin film was cut into small pieces and transferred onto the fiber end. To enhance the effect of saturable absorption, 5 pieces of WS$_2$-PVA thin film were transferred and stacked on the fiber end as WS$_2$-PVA SA, shown in Figure 1(c). The total thickness of WS$_2$-PVA SA is ~100 μm. To confirm the incorporation of WS$_2$ nanosheets into the PVA thin film, the transmission spectra of WS$_2$-PVA thin film and pure PVA thin film were characterized with a spectrometer (PerkinElmer Lambda 750 instrument), shown in Figure 1(d). The pure PVA thin film has a transmission of ~98% near 1550 nm and the WS$_2$-PVA thin film has a transmission of ~96% near 1550 nm. The dip near 632 nm (1.96 eV) in the transmission spectrum of WS$_2$-PVA thin film is a typical fingerprint of WS$_2$ nanosheets due to the direct bandgap transition [1]. Raman spectroscopy was also employed to further confirm the incorporation of WS$_2$. A Raman spectroscopy system (Renishaw invia) with an excitation wavelength of 488 nm was utilized to investigate the atomic structural arrangement of WS$_2$, as shown in Figure 1(e). The in-plane vibrational mode $E^1_{2g}$ at 355.9 cm$^{-1}$ and the out-of-plane vibrational mode $A_{1g}$ at 420 cm$^{-1}$ can be clearly observed. The frequency difference between the $E^1_{2g}$ mode and $A_{1g}$ mode reveals the number of layers of WS$_2$ nanosheets [7]. Therefore, the frequency difference of ~64.1 cm$^{-1}$ in this work implied the number of layers to be ~3. For comparison, the Raman spectrum of the WS$_2$ dispersions without PVA was also measured. It can be seen that the Raman spectra of WS$_2$ dispersions and WS$_2$-PVA thin film were nearly identical, which indicates that the WS$_2$ nanosheets were embedded in the PVA thin film with nearly intact atomic structures. This finding suggests that PVA is a suitable host for WS$_2$ nanosheets and filmy WS$_2$-PVA SA provides significant flexibility in the practical applications due to its compactness compared with aqueous dispersions.

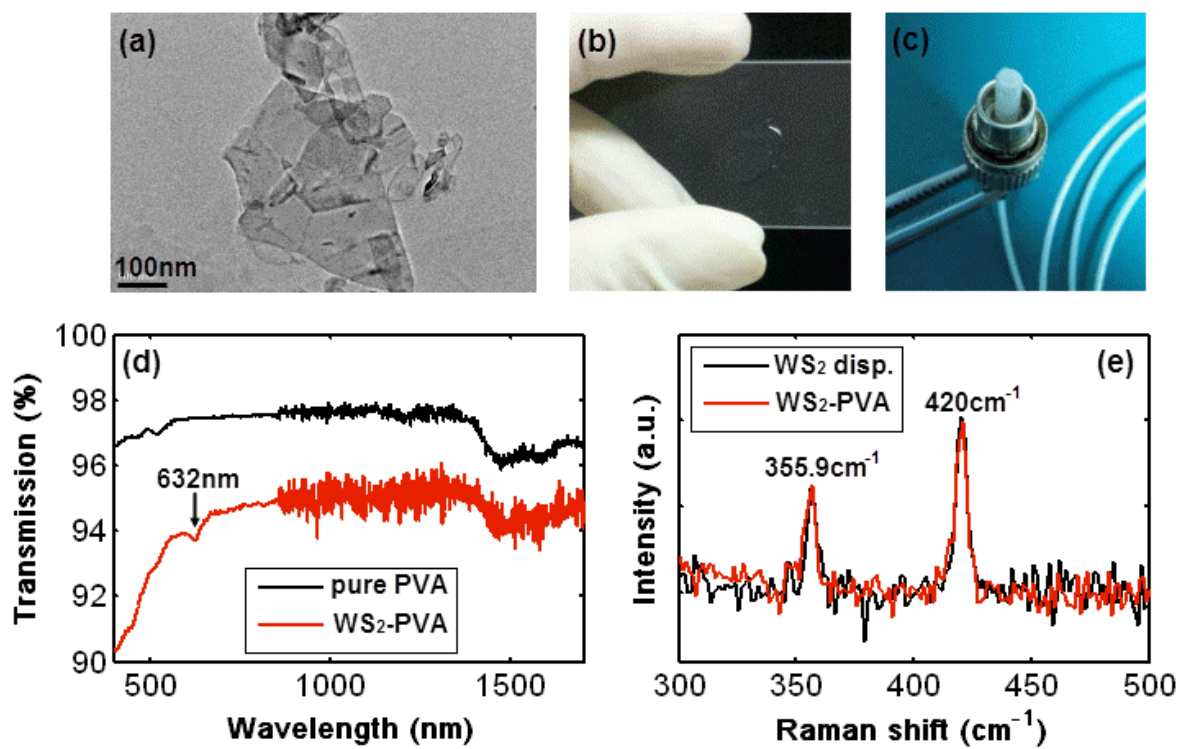

**Figure 1** (a) TEM image of WS$_2$ nanosheets, (b) WS$_2$-PVA thin film on a glass plate, (c) WS$_2$-PVA SA on fiber end, (d) transmission spectra of WS$_2$-PVA thin film and pure PVA thin film and (e) Raman spectra of WS$_2$ dispersions without PVA and WS$_2$-PVA thin film.

To investigate the saturable absorption of the prepared WS$_2$-PVA SA, the nonlinear transmission property was characterized at the telecommunication wavelength using a standard 2-arm transmission measurement scheme, shown in Figure 2(a). A commercial mode-locked laser was applied as a pulsed source. The laser had a center wavelength of 1560 nm, repetition rate of 37 MHz, pulse width of ~510 fs and maximum output power of ~20 mW. The output of the laser was split into two arms with the upper arm for the power-dependent transmission measurement of $WS_2$-PVA SA and the lower arm for reference. The measurement results were shown in Figure 2(b). The measurement was performed first by increasing the input power (or intensity) and then by decreasing the input power. The nearly unchanged results from the two measurements confirmed the existence of saturable absorption in $WS_2$-PVA SA. The modulation depth of the $WS_2$-PVA SA is ~2.96% with saturation intensity of ~362 MW/cm$^2$. The non-saturable loss is ~30.9%. The non-saturable loss is due to the scattering and absorption of the $WS_2$ flakes and PVA thin film, and the coupling loss between two fiber connectors.

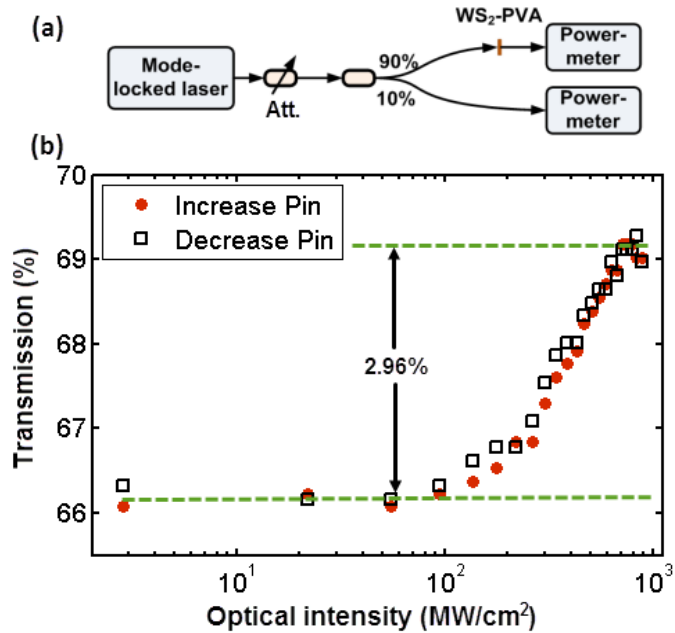

**Figure 2** (a) Measurement setup of the nonlinear transmission of $WS_2$-PVA SA and (b) measured saturable absorption with a modulation depth of 2.96%.

It should be mentioned that although the direct bandgap of monolayer $WS_2$ is ~2.0 eV (~630 nm) and the indirect bandgap is 1.4 eV (886 nm) [40], the saturable absorption was observed near 1550 nm corresponding to a photon energy of 0.8 eV which is even smaller than the minimum bandgap. Similar to the explanation for $MoS_2$ in the introduction, this confliction may be attributed to the existence of the defects in the material. Since $WS_2$ has material property similar to $MoS_2$ and the bandgap of $MoS_2$ can be experimentally reduced to 0.26 eV (4.7 μm) by introducing defects [34], it is reasonable to deduce that the defects in $WS_2$, which is inevitable in the preparation of the materials, also reduce the bandgap and expand the waveband for the existence of saturable absorption.

## 2.2. Passively mode-locked laser

Passively mode-locked fiber lasers with femtosecond output pulses attract intense interest due to its wide applications in communications, sensing and frequency metrology. We first demonstrated the $WS_2$-PVA SA as a mode locker for a passively mode-locked fiber laser near 1550 nm. The laser setup is shown in Figure 3. The detailed description of the laser design can be referred to the Methods section. The $WS_2$-PVA SA was embedded between two fiber connectors in the cavity.

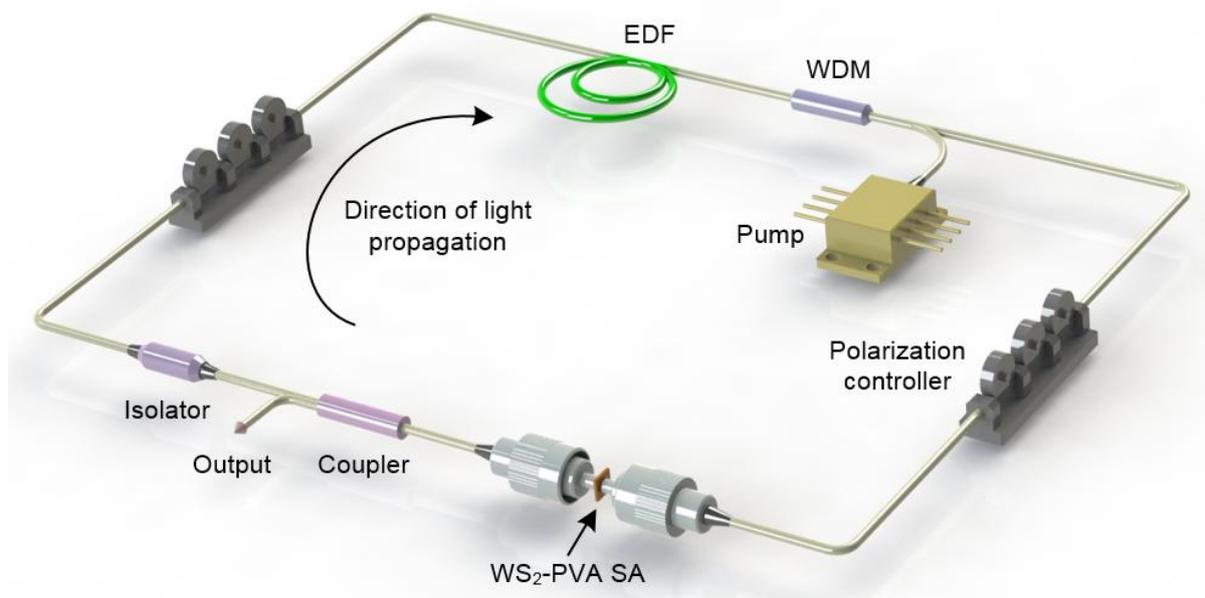

**Figure 3** Experimental setup of the fiber laser with $WS_2$-PVA SA. EDF: Erbium-doped fiber; WDM: wavelength-division multiplexer.

The laser output properties are summarized in Figure 4. The laser self-started with harmonic mode locking at a pump power near 350 mW. Fundamental mode locking was obtained when the pump power was decreased to 260 mW and then the fundamental mode locking can be sustained at the pump power ranging from ~180 mW to 310 mW. At the pump power of 280 mW and fundamental mode locking operation, the laser had an optical spectrum centered at 1572 nm with a 3-dB bandwidth of 5.2 nm, shown in Figure 4(a). Clear Kelly sidebands can be observed which confirmed the soliton mode locking operation of the laser. The autocorrelation trace in Figure 4(b) had a width of 919 fs corresponding to a pulse width of 595 fs assuming $sech^2$ profile. To confirm the stable mode locking operation, a 10-GHz photodetector and a 2.5-GHz real-time oscilloscope were employed to investigate the pulse train in time domain, shown in Figure 4(c). Very uniform pulse amplitude can be observed which indicates the laser was operating in the continuous wave mode locking mode. The laser output power with respect to the pump power was shown in Figure 4(d) and the regions of fundamental and harmonic mode locking were also denoted. RF spectra were also measured to evaluate the spectral quality of the mode locking operation, shown in Figure 5. In a span of 400 kHz with resolution bandwidth (RBW) of 10 Hz, the extinction ratio is 75 dB. And in a span of 500 MHz with RBW of 10 kHz, no unwanted sidebands or spurious peaks were observed which indicates the high spectral purity of the mode locking operation.

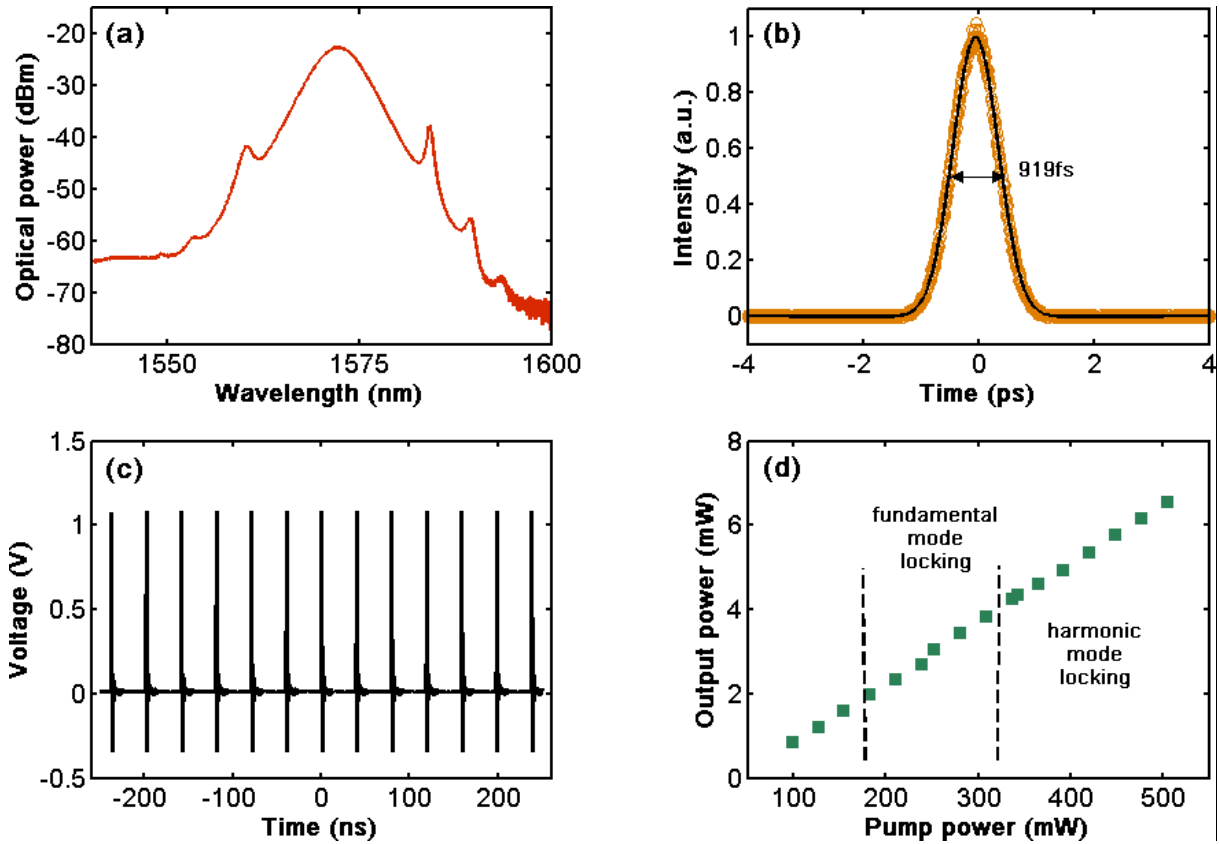

**Figure 4** (a) Optical spectrum, (b) autocorrelation trace, (c) oscilloscope trace and (d) output power with respect to pump power of the mode-locked laser based on $WS_2$-PVA SA.

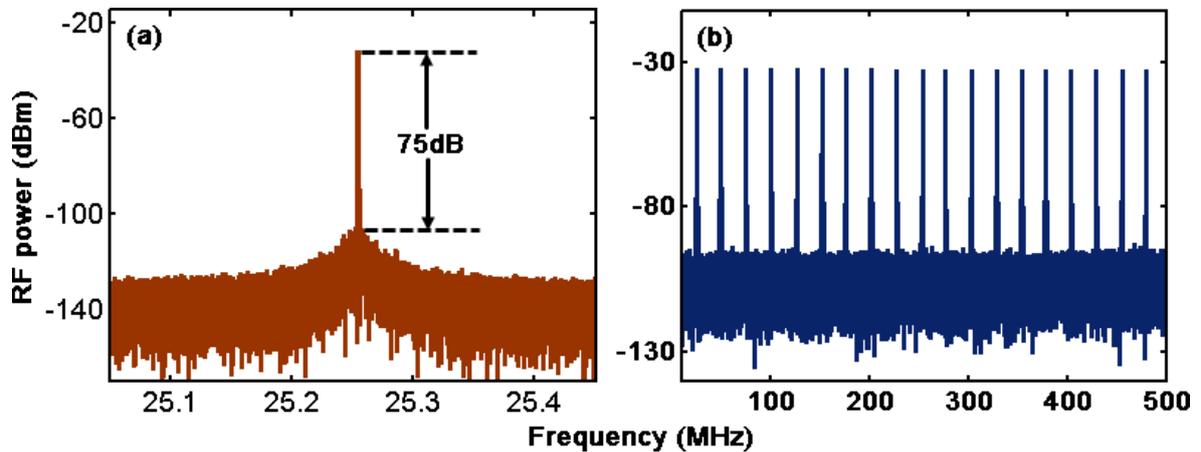

**Figure 5** RF spectra of the laser for (a) a span of 400 kHz with resolution bandwidth of 10 Hz and (b) a span of 500 MHz with resolution bandwidth of 10 kHz.

### 2.3. Q-switched laser

Besides mode locking, Q-switching operation is another important operation mode in the fiber lasers, which provides tunable repetition rate and relatively high pulse energy. By adjusting the birefringence in the laser cavity with polarization controllers, Q-switching operation can also be obtained in the same laser cavity incorporated with $WS_2$-PVA SA. Figure 6 summarizes the output properties of the Q-switching operation. The Q-switching operation existed at the pump power ranging from 252 mW to 364 mW. Out of this range, continuous wave operation dominated. In the Q-switching operation, the laser had a center wavelength of 1570 nm, shown in Figure 6(a). Oscilloscope trace in Figure 6(b) confirmed the stable pulsed operation of the laser and the repetition rate was dependent on the pump power varying from 90 kHz to 125 kHz, shown in Figure 6(c). The relation between the laser output power and pump power was given in Figure 6(d) with the region of Q-switching operation denoted. The maximum pulse energy was 46.3 nJ at the pump power of 280 mW.

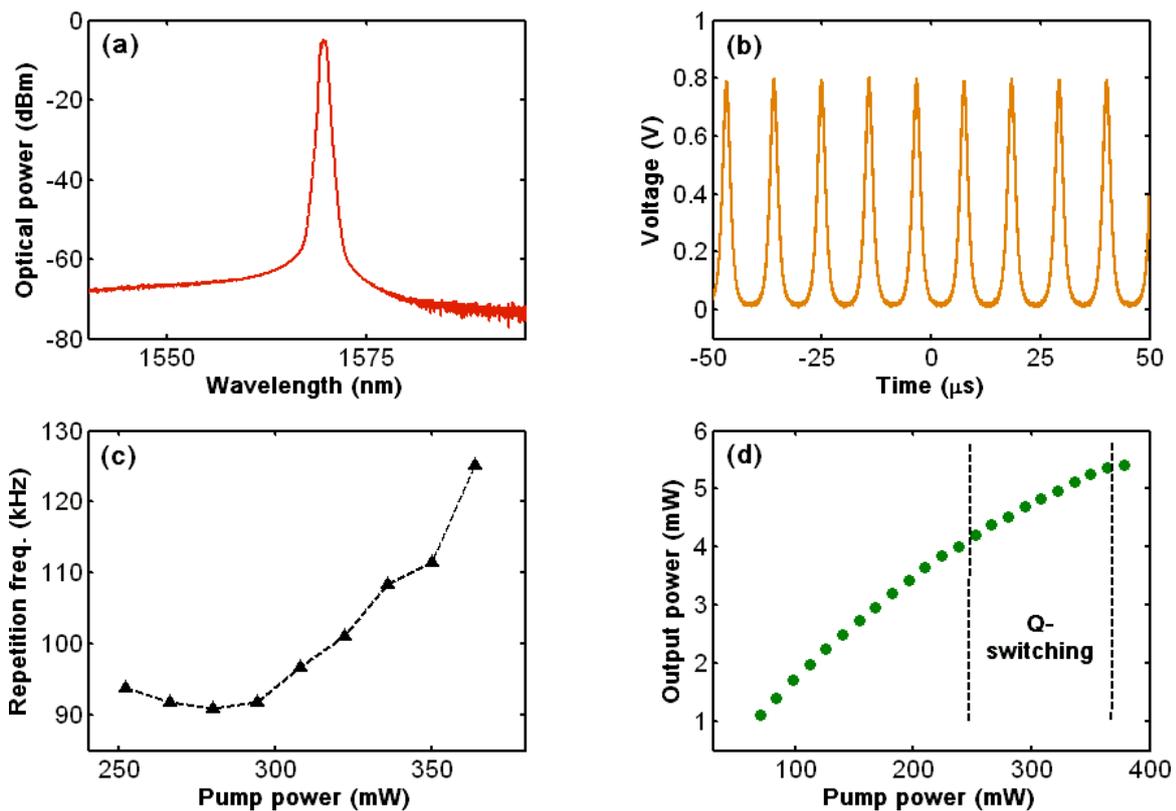

**Figure 6** (a) Optical spectrum, (b) oscilloscope trace, (c) repetition rate and (d) output power with respect to pump power of the Q-switched laser based on $WS_2$-PVA SA.

## 3. Conclusions

WS$_2$-PVA saturable absorber was prepared with a modulation depth of 2.96% and a saturation intensity of 362 MW/cm$^2$ for ultrafast photonic applications. Both mode locking and Q-switching operations near 1550 nm were demonstrated in a fiber laser incorporated with WS$_2$-PVA SA. In the mode locking operation, femtosecond pulse width of 595 fs and high RF spectral purity of 75 dB extinction ratio were obtained whereas in the Q-switching operation tunable repetition rate from 90 kHz to 125 kHz and pulse energy with tens of nJ were obtained. These findings demonstrate the potential of WS$_2$-PVA SA as a promising mode locker and Q switcher for the ultrafast photonic applications. Moreover, the thin-film form provides significant flexibility in the practical use due to its compactness and easy fabrication.

## 4. Methods

### 4.1. Preparation of WS$_2$ dispersions

The high-quality WS$_2$ dispersions were prepared by liquid-phase exfoliation method using ionic surfactant SC (Sigma-Aldrich) as a stabilizer [3]. Typically, 5 mg/ml of WS$_2$ powders (Sigma-Aldrich) were dispersed in 1.5 mg/ml SC aqueous solution and sonicated for 1 hour using a horn probe sonic tip (VibraCell CVX; 750 W) with 38% output power. The dispersions were then centrifuged at 3000 rpm for 90 minutes to remove the unexfoliated ones. The top 2/3 of the dispersions was collected by pipette.

### 4.2. Laser design and measurement

The laser cavity consisted of ~1 m Erbium-doped fiber with anomalous dispersion and ~7.2 m standard single mode fiber. The net cavity dispersion is estimated to be ~-0.18 ps$^2$ and the laser was operating in the soliton mode locking regime. Pump light at 976 nm was injected into the cavity via a wavelength division multiplexer (WDM). Two polarization controllers were used to adjust the cavity birefringence. An isolator was incorporated to guarantee the single-direction operation and a 90:10 coupler extracted 10% intra-cavity power for output. The WS$_2$-PVA SA was embedded between two FC/PC fiber connectors. The optical spectra were measured by an optical spectrum analyzer (Yokogawa AQ6370C). The autocorrelation trace was measured by an autocorrelator (Femtochrome 103XL). The oscilloscope traces were measured by a 10-GHz photodetector (EOT 3500F) and a 2.5-GHz real-time oscilloscope (Agilent DSO9254A). The RF spectra were measured by a 44-GHz RF spectrum analyzer (Agilent N9010A EXA).

**Acknowledgements**

This work is partially supported by the Shanghai Yangfan Program (No. 14YF1401600), the State Key Lab Project of Shanghai Jiao Tong University (No. GKZD030033), NSFC (No. 61178007, No. 51302285), STCSM (Nano Project No. 11nm0502400, Shanghai Pujiang Program 12PJ1409400), the External Cooperation Program of BIC, CAS (No. 181231KYSB20130007). J. W. thanks the National 10000-Talent Program and CAS 100-Talent Program for financial support.

**Author contributions**

K. W. designed and performed the experiments. X. Z. and J. W. fabricated the $WS_2$ dispersions and performed the material characterization. X. L. assisted the experiments. J. C. provided helpful discussions. All the authors contributed to the analysis of the experimental data and the preparation of the manuscript.

The authors declare no competing financial interests.

**Supplementary Information**

NA

**Figure Caption**

Figure 1 (a) TEM image of $WS_2$ nanosheets, (b) $WS_2$-PVA thin film on a glass plate, (c) $WS_2$-PVA SA on fiber end, (d) transmission spectra of $WS_2$-PVA thin film and pure PVA thin film and (e) Raman spectra of $WS_2$ dispersions without PVA and $WS_2$-PVA thin film.

Figure 2 (a) Measurement setup of the nonlinear transmission of $WS_2$-PVA SA and (b) measured saturable absorption with a modulation depth of 2.96%.

Figure 3 Experimental setup of the fiber laser with $WS_2$-PVA SA. EDF: Erbium-doped fiber; WDM: wavelength-division multiplexer.

Figure 4 (a) Optical spectrum, (b) autocorrelation trace, (c) oscilloscope trace and (d) output power with respect to pump power of the mode-locked laser based on $WS_2$-PVA SA.

Figure 5 RF spectra of the laser for (a) a span of 400 kHz with resolution bandwidth of 10 Hz and (b) a span of 500 MHz with resolution bandwidth of 10 kHz.

Figure 6 (a) Optical spectrum, (b) oscilloscope trace, (c) repetition rate and (d) output power with respect to pump power of the Q-switched laser based on $WS_2$-PVA SA.